# The Nitrogen concentration effect on Ce doped SiO$_x$N$_y$ emission: towards optimized Ce$^{3+}$ for DEL applications

F. Ehré[1], C. Labbé[1*], C. Dufour[1], W. M. Jadwisienczak[2], J. Weimmerskirch-Aubatin[1], X. Portier[1], J.-L. Doualan[1], J. Cardin[1], A. L. Richard[3], D. C. Ingram[3], C. Labrugère[4], F. Gourbilleau[1]

[1] CIMAP, Normandie Univ, ENSICAEN, UNICAEN, CEA, CNRS, 14050 Caen, France

[2] School of Electrical Engineering and Computer Science, Ohio University, Stocker Center, Athens, OH 45701, USA

[3] Department of Physics and Astronomy, Ohio University, Athens, OH 45701, USA

[4] PLACAMAT, UMS 3626, CNRS-Université Bordeaux, 87 avenue Albert SCHWEITZER, 33600 Pessac, France

* corresponding author: christophe.labbe@ensicaen.fr

## Abstract

Ce-doped SiO$_x$N$_y$ films are deposited by magnetron reactive sputtering from a CeO$_2$ target under nitrogen reactive gas atmosphere. Visible photoluminescence measurements regarding the nitrogen gas flow reveal a large emission band centered at 450 nm for a sample deposited under a 2 sccm flow. A special attention is paid to the origin of such an emission at high nitrogen concentration. Different emitting centers are suggested in Ce doped SiO$_x$N$_y$ films (*e.g.* band tails, CeO$_2$, Ce clusters, Ce$^{3+}$ ions), with different activation scenarios to explain the luminescence. X-ray photoelectron spectroscopy (XPS) reveals the exclusive presence of Ce$^{3+}$ ions whatever the nitrogen or Ce concentrations, while transmission electron microscopy (TEM) shows no clusters or silicates upon high temperature annealing. With the help of photoluminescence excitation spectroscopy (PLE), a wide excitation range from 250 nm up to 400 nm is revealed and various excitations of Ce$^{3+}$ ions are proposed involving direct or indirect mechanisms. Nitrogen concentration plays an important role on Ce$^{3+}$ emission by modifying Ce surroundings, reducing the Si phase volume in SiO$_x$N$_y$ and causing a nephelauxetic effect. Taking into account the optimized nitrogen growth parameters, the Ce concentration is analyzed as new parameter. Under UV excitation, a strong emission is visible to the naked eye with high Ce$^{3+}$ concentration (6 at. %). No saturation of the photoluminescence intensity is observed, confirming again the lack of Ce cluster or silicate phase formation due do the nitrogen presence.

## Introduction

Rare earth (RE) doping of silicon host matrices has been widely studied as a method to add light emitting sources to integrated optoelectronics devices using the RE inter-4f transitions that offer multiple wavelength emission. Typically, those transitions, forbidden in the case of a free ion, are partially allowed by the crystal field and result in a very low absorption cross section inducing a non-efficient excitation and thus limiting the applications.[1] $Ce^{3+}$ ion is different due to a 5d-4f transition involved resulting in a large absorption cross section ($10^{-19}$ $cm^{-2}$) compared to the other RE ions ($10^{-21}$ $cm^{-2}$). An intense broad luminescence band from these $Ce^{3+}$ ions is observed in a wide spectral range from 380 nm up to 600 nm under various excitations ranging from 300 nm to 400 nm due to the strong oscillator strengths[2] and many energy levels of the 5d band involved.[3] Furthermore, in the case of $Ce^{3+}$ ion having only one electron within the 5d orbital, which participate in the formation of the chemical bonding, this 5d band is strongly dependent on the local environment, resulting in a large Stockes shift depending on the host matrix composition.[2–4]

These different major properties of $Ce^{3+}$ ion (*e.g.* intense luminescence, broad absorption/emission band, dependence of the excitation/emission wavelength on the crystal structure, *etc.*) have been developed for many applications such as phosphors for cathodoluminescence light sources,[5] scintillators for elementary particles,[6] or detectors for ionizing radiation.[7] Two other applications became extensively explored, namely quantum cutting for solar cells and UV emitting devices.[8,9] For the former, the quantum cutting effect is created with help of $Ce^{3+}$-$Yb^{3+}$ co-doping in glasses,[9,10] phosphor[8,11] or bulk crystal matrices.[12] Although much effort has been spent on depositing such matrices on commercial silicon solar cells,[11,13] their properties are still not silicon technology compatible. That is why recent researchers have turned to investigate host matrices easily integrable on silicon such as silicon oxide.[14,15] Regarding the Ce doped devices devoted to light emitting devices on silicon, the feasibility of a $CeO_2$ and $Ce_2Si_2O_7$ LEDs has been demonstrated,[16,17] while a Ce-doped $SiO_x$ matrix shows a strong emission of the $Ce^{3+}$ ion.[18,19]

Unfortunately, it exists some major drawbacks hampering such applications. In particular, $Ce^{3+}$ and other $RE^{3+}$ dopants have a low solubility limit. For instance, this leads to Ce clustering in $SiO_x$ after annealing at 900°C.[19] In addition, silica based LEDs present a short durability after several hot carrier injection cycles under high applied voltage because of its large 9 eV bandgap.[20] To overcome the above mentioned drawbacks, the silicon compatible $Si_3N_4$ host matrix is also studied as an alternative approach. Indeed, the $Si_3N_4$ has much higher RE ions solubility in comparison to $SiO_x$ counterpart, thus inhibiting RE ions clustering.[21,22] Moreover, its lower bandgap (4-5 eV)[23] can improve both electrical conductivity and the onset voltage for the electroluminescence signal.[24,25] Since the presence of oxygen is crucial for a $Ce^{3+}$ emission,[26] one possibility is to combine advantages of $SiO_2$ (*i.e.* presence of oxygen) and $Si_3N_4$ (*i.e.* lower band gap and higher solubility). In that regards, the $SiO_xN_Y$ matrix retains the advantages of both matrices as it was demonstrated for a LED based on $Ce^{3+}$-doped $SiO_xN_y$.[20]

In this context, it is interesting to investigate how the Ce element is incorporated into $SiO_xN_y$. In the past, Ce doping from metallic target was reported.[18,19,27] However, metal targets are very sensitive to surface oxidation due to the electronic affinity between Ce and O elements thus their use requires a very high temperature (~1400 °C). Above all, the Ce metallic target contains more Ce element than the $CeO_2$ target. Such high purity can result in inhomogeneous Ce dopants deposition and favor the growth of Ce clusters at high annealing temperature. Thus, an alternative solution is to sputter from a $CeO_2$ target, with both benefits of a chemical stability and a much lower target temperature required during the deposition process. This approach can generate $Ce^{4+}$ ions which are optically inactive due to their electronic configuration. Nevertheless, numerous studies have demonstrated the efficient emission of $Ce^{3+}$ in samples deposited from $CeO_2$ targets.[15–17] Unfortunately, depositing from $CeO_2$ target requires a post-growth annealing step at high temperature to form optically active $Ce^{3+}$ ions which could generate Ce clusters due to the high thermal budget. Considering the above mentioned arguments, we believe that doping the $SiO_xN_y$ host matrix with Ce appears to be an interesting approach for achieving efficient cerium emission.

In this work, a $CeO_2$ target was used to incorporate $Ce^{3+}$ ions in a $SiO_xN_y$ matrix by reactive magnetron sputtering which is easily compatible with Si processing technology. Despite the use of a $CeO_2$ target which contains optically inactive $Ce^{4+}$ ions, a specific $Ce^{3+}$ ion emission is obtained with an annealing temperature ($T_A$) of 700 °C. XPS, Rutherford backscattering spectroscopy (RBS) and spectroscopic ellipsometry measurements on Ce-doped $SiO_xN_y$ show the influence of nitrogen composition on the host matrix structure. Photoluminescence (PL) and photoluminescence excitation (PLE) experiments were carried out to study in the first step, the $Ce^{3+}$ ion excitation mechanisms, specifically the link between nitrogen content and the Ce excitation. A second part is focused on the $Ce^{3+}$ emission in function of the Ce concentration at high nitrogen flow, in order to optimize the PL intensity.

**Experiment section**

Ce doped $SiO_xN_y$ thin layers were deposited at room temperature by magnetron reactive sputtering on p-type doped 250 µm thick 2" (100 oriented) Silicon wafer. The working pressure was fixed at 3 mTorr and the argon flow was set at 8 sccm. Two sets of samples have been deposited using $CeO_2$ and Si targets. A set of fours Ce doped samples was grown under variable $N_2$ flux and a second set of four samples with different cerium doping, including one undoped as reference. In the case of this specific reference sample, the $CeO_2$ target was replaced by a $SiO_2$ target to ensure oxygen incorporation in the matrix. No study has been performed above 2 sccm $N_2$ flux because the deposition chamber atmosphere was not enough plasmogenic, stopping the plasma and then the deposition process. The atomic compositions as well as the refractive index measurements have confirmed the same stoichiometry between undoped and doped layers. Films were then annealed at different temperatures ($T_A$) from 600 °C to 1000 °C using Classic Thermal Annealing (CTA) setup for 1 h under flowing $N_2$.

Structural analysis were performed using a Fourier transform infrared spectrophotometer (FTIR), Thermo Nicolet Nexus 750 II working in the 400-4000 $cm^{-1}$ range with a 5 $cm^{-1}$ resolution at Brewster angle, while RBS measurements were carried out at the *Edwards Accelerator Laboratory* of Ohio University using a 4.5 MV tandem accelerator. The film thicknesses and refractive indexes were obtained with help of UVISEL VIS-FGMS Ellipsometer using an incident angle of 70°. The experimental data were recorded in the 1.5 - 6 eV range with a 0.01 eV resolution and fitted using *DeltaPsi2* software in order to obtain the thickness and the refractive index values. Except if mentioned otherwise, this later is given at 1.95 eV (636 nm).

For the PL spectroscopy, a Lot-Oriel 1 kW Xenon lamp connected to an OMNI300 monochromator resulting in a 20 nm wavelength wide excitation band was used to excite the sample between 325 nm and 370 nm. The detection system consisted in a photomultiplier tube R5108, (Hamamatsu Photonics) attached to a 300 OMNI monochromator (Gilden Photonics) and a SR830 lock-in amplifier referenced at the excitation light beam chopped frequency. A ThermoFisher Scientific K-Alpha spectrometer was used for surface and in-depth XPS analysis. The monochromatic Al-$K\alpha$ source ($h\nu$ = 1486.6 eV) was used with a spot size of 200 µm in diameter. The spectra in full energy range (0 - 1150 eV) were obtained with a constant pass energy of 200 eV whereas high resolution spectra were obtained with constant pass energy of 40 eV, respectively. Depth profile analysis was obtained through $Ar^+$ sputtering using 500 eV low mode and 500 microns raster width. The estimated sputtering rate was 0.2 nm/s resulting in the etching time between 0 s and 960 s needed to reach the substrate. The XPS measurements were performed at different depths; however only spectra collected at the film center are presented here. Note that carbon t was detected on the surface coming from natural pollution and was not taken into account. Electron microscopy analysis was performed with a JEOL 2010F electron transmission microscope (TEM) with a Field Emission Gun (FEG) electron source operated at 200 kV and whose resolution was 2 Å. The digitalized images were processed by the Gatan's *DigitalMicrograph* software. The TEM was equipped with an energy dispersive X-ray spectrometer (EDX-EDAX setup) for chemical analyses. The EDX profiles were performed using a 1 nm probe for 30 points over a distance of 100 nm. The cross sectional samples were prepared using a Focused Ion Beam (FIB) *Helios DualBeam nanolab660* system. To protect the film surface from the gallium ions beam induced damage, a protective carbon layer and two platinum layers were deposited on the sample's surface.

## Results and Discussions

In order to study the behavior of the photoluminescence (PL) of Ce-doped $SiO_xN_y$ films, a set of samples has been deposited with the same RF power density on silicon (4.3 W.cm$^{-2}$) and $CeO_2$ (0.45 W.cm$^{-2}$) targets, under different nitrogen fluxes. The emission spectra as well as the structural analysis are presented and discussed below.

**Photoluminescence and Photoluminescence excitation spectroscopy**

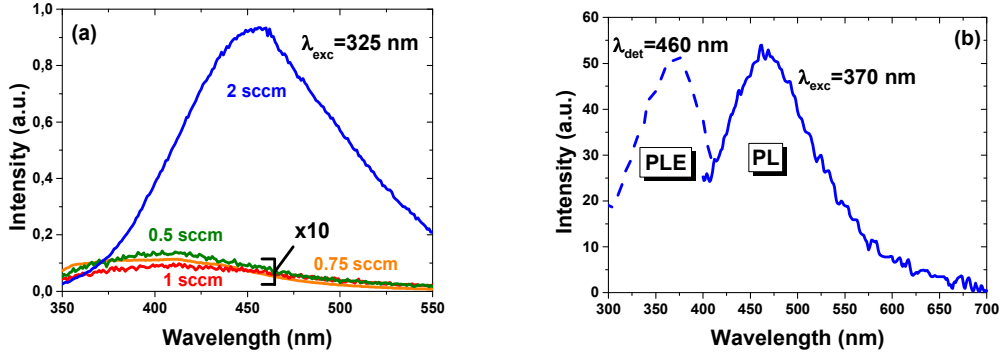

*Figure 1. (a) PL spectra of Ce-doped SiO$_x$N$_y$ samples grown with different nitrogen fluxes and annealed at 700°C under 325 nm excitation normalized by the film thicknesses, (b) PL (solid line) and PLE (dashed line) spectra of Ce-doped SiO$_x$N$_y$ deposited with a 2 sccm nitrogen flux.*

Figure 1a displays photoluminescence (PL) spectra of the four Ce-doped SiO$_x$N$_y$ films grown with four different N$_2$ flow rates (0.5, 0.75, 1 and 2 sccm) and using 325 nm excitation. Because no emission was observed (not shown) for as deposited layers, all the samples Figure 1a were annealed at 700 °C. A low intensity and broad PL band centered at ~400 nm is observed for the samples grown with the three lowest N$_2$ fluxes (0.5 sccm, 0.75 and 1 sccm). Such a broad emission band is the signature of band tails (BT) originating from localized states associated with defects such as dangling bonds in the host matrix.[25,28–30] In the past, various studies have reported the role played by such localized states structuring the BT acting as sensitizers for RE emission.[22] Indeed, this observed BT peak blue shifts greatly with the excitation wavelength decrease ($\lambda_{exc}$= 500 nm, BT peak centered at 550 nm while for $\lambda_{exc}$=300 nm, it is peaking at 400 nm).[28] Detection of the almost same peaks for the three samples grown at low N$_2$ fluxes shows that the Ce emission, or at least other emitter centers, are not efficient and only the BT emission appears.

For the N$_2$ flow rate of 2 sccm a broad and intense PL band peak position red-shifted to 450 nm. The observed PL intensity is 90 times higher than that of BT from samples deposited at lower N$_2$ flow rates. This PL emission could originate from different emitting centers including: *(i)* states localized in the BT, *(ii)* Ce$^{3+}$ ions assisted by oxygen vacancies in CeO$_2$ involving Ce$^{4+}$ ions following the Kroger-Vink notation: $4Ce^{4+} + O^{2-} => 2Ce^{4+} + 2Ce^{3+} + V_o + \frac{1}{2}O_2$ with $V_0$ referring to oxygen vacancy,[16,31] *(iii)* formation of Ce oxide single crystals (Ce$_6$O$_{11}$) when taking into consideration the post-growth low T$_A$ treatment (< 800°C),[31] or *(iv)* "isolated" Ce$^{3+}$ ions.

Concerning the case *(i)*, the increase of the N$_2$ flux could provide an increase of the localized states density leading to the rise of the PL intensity. However, the 90 nm red shift of the PL peak maximum, especially if the excitation wavelength is unchanged, excludes the origin of the BT emission. The emitter centers could appear in the CeO$_2$ films (*case (ii)*) or in the Ce$_6$O$_{11}$ crystallized phase (*case (iii)*). These points will be discussed separately later. To investigate the Ce$^{3+}$ emission possibility (*case (iv)*), the PLE spectra of Ce-doped SiO$_x$N$_y$ matrix deposited with a 2 sccm N$_2$ flux was compared to its PL spectrum counterpart (Figure 1b). The PLE spectrum shows a wide excitation peak (FWHM ~75 nm) with a maximum at 370 nm. The observed PLE band is characteristic of the 5d level of Ce$^{3+}$ ion due to the transition from the $^2F_{5/2}$ ground level.[19] In the present study, Ce$^{3+}$ ions are directly excited from a 370 nm excitation wavelength what is consistent with previous studies. For example, for Ce-doped phosphor, the PLE peak was centered at 357 nm and its FWHM was ~75 nm,[32] while for Ce-doped $SiO_{1.5}$, a similar matrix to the one studied here, a PLE peak was centered at 300 nm with a FWHM of almost 50 nm.[19] Moreover, an important Stokes shift of 97 nm (5600 cm$^{-1}$) was found, which is similar to the one observed in $Ce^{3+}$ ion doped inorganic matrices.[33]

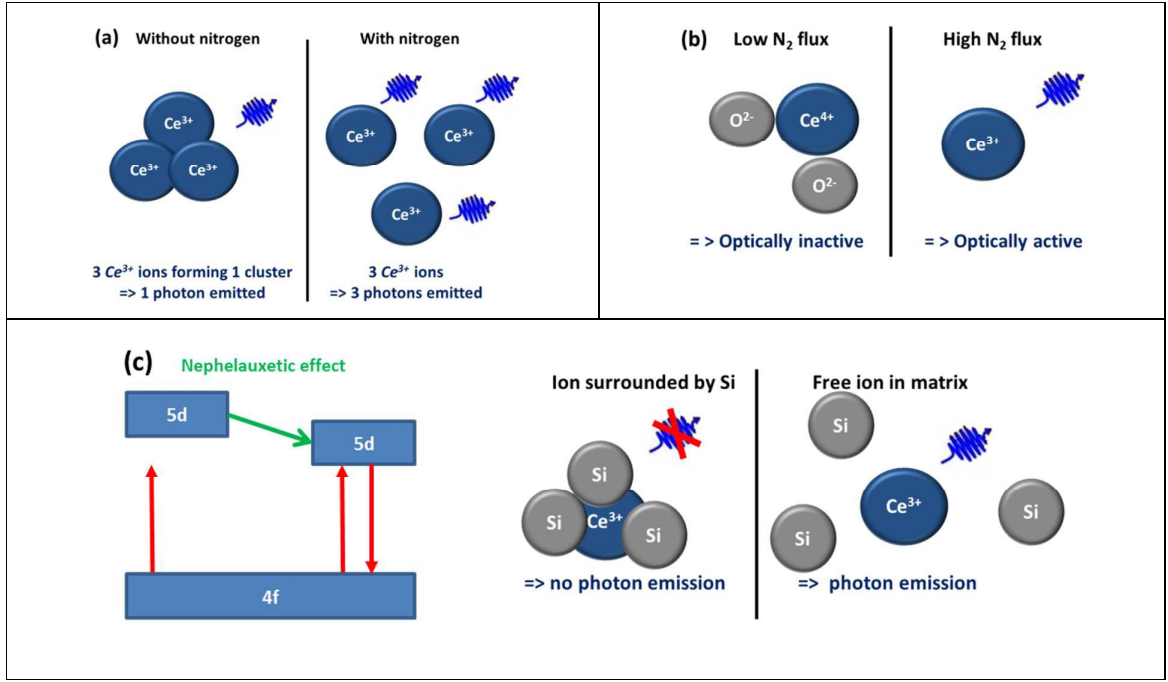

*Figure 2. Schematic representation of possible scenarios illustrating the formation of Ce optical centers in $SiO_xN_y$ matrix deposited with reactive $N_2$ flux at 2 sccm (a) α: cluster formation reducing $Ce^{3+}$ ion emission intensity (b) β: Ce oxidation state depending on nitrogen content (c) γ: nitrogen effect on Ce surrounding media*

Assuming that the increase of PL intensity is not due to an important rise in the Ce concentration (see below *Section RBS*), the emitting centers are activated by an higher reactive $N_2$ flux (between 1 sccm to 2 sccm) during growth. At this stage, we shall also consider the different plausible scenarios of an activation of such Ce emitting centers affected by the N incorporation as schematically illustrated in Figure 2.

*Case (α)*: Nitrogen is known to reduce the RE clustering effect.[34] The N incorporation should prevent Ce clusters formation and favors the $Ce^{3+}$ emission (see Figure 2a).

*Case (β)*: Especially in presence of an oxygen excess in the layer, oxide defects are known to favor the $Ce^{4+}$ formation. Thermal treatment reduces the number of oxide defects, helping a valence conversion of Ce from (IV) to (III).[35,36] The introduction of N could also change the local distribution of oxygen and facilitate the conversion from optically inactive $Ce^{4+}$ ion to optically active $Ce^{3+}$ ion as shown in Figure 2b.[36]

*Case (γ)*: The strong dependence of the Ce 5d band on the surrounding host structure could be involved. In this scenario, two indiscernible mechanisms could apply at the same time.

First, the increase of the N content in the matrix changes the local environment of the Ce ion, inducing a broadening of the electronic cloud, also known as the nephelauxetic effect,[37] which could lower the energy necessary for an optical activation. Second, it has been reported that the Ce ions emission could be killed when surrounded by Si atoms.[19] The higher $N_2$ flux could reduce the number of Si neighbors around Ce ion and create a favorable emission environment. Those two last mechanisms resulting from the different surroundings of the Ce ion are schematized in Figure 2c.

In the following, structural investigations are carried out to answer these different assumptions concerning the nature and the way how such emitting centers are excited in the Ce-doped $SiO_xN_y$ sample produced under a 2 sccm $N_2$ flow.

## Microstructural analysis – TEM observations

TEM, HRTEM and Fast Fourier Transform (FFT) images from Ce-doped $SiO_xN_y$ samples deposited with 0.75 sccm and 2 sccm $N_2$ fluxes and annealed at 700°C and 900°C are presented Figure 3.

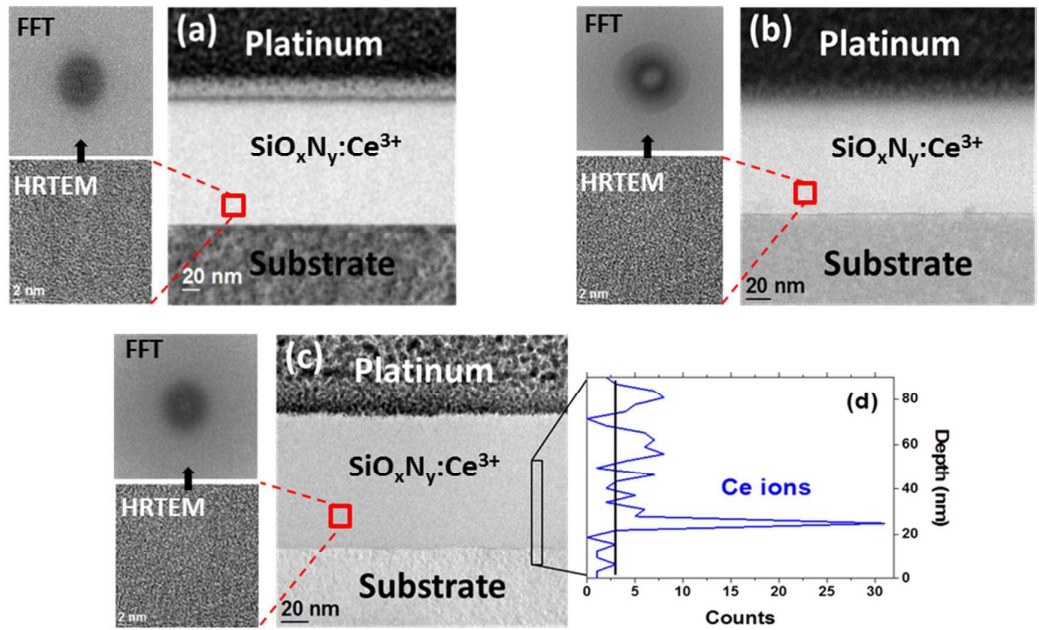

*Figure 3. TEM pictures of Ce-doped $SiO_xN_y$ deposited with (a) 0.75 sccm $N_2$ flux upon annealing at 700 °C. (b) and (c) 2 sccm $N_2$ flux upon annealing at 900 °C and 700 °C, respectively. Panels (a), (b) and (c) contains HRTEM images with the corresponding FTT analysis (d) Ce ions profile distribution from (c) extracted from EDX measurement. The vertical line indicates the maximum EDX background signal not related to Ce ions.*

In each case, a bright field image is shown and an HRTEM image from a small region of the film is enlarged at the bottom left handside of the image. At the top left handside, FFT of the HRTEM image is depicted. Both 700°C and 900°C annealed films display amorphous and homogeneous films without phase separation whatever the $N_2$ flux. Note that for as-grown samples deposited with 0.75 sccm and 2 sccm, the same amorphous film structure has been found (not shown). Concerning the SiO$_x$ host matrices with 1<x<2, some previous studies have been reported with respect to different RE doping and post-growth $T_A$. For example, in SiO$_2$, Tb or Eu-oxide clusters have been identified in the middle of the film after annealing at 900 °C,[38] while in Er-doped silicon rich silicon oxide (SRSO), Er-clusters appears after annealing at 800 °C, limiting for both cases the luminescence efficiency.[39] Concerning the Ce doping, the same clustering effect has been observed at $T_A$ =900°C in SiO$_{1.5}$.[19,35,40] However, in the present study, all the HRTEM images do not show lattice fringes and very diffuse rings characteristic of amorphous materials are observed in the FFT images. In other words, no trace of Ce clustering effect has been detected in any film annealed at 700 °C or 900 °C. This result confirms that the introduction of nitrogen in silicon-based matrix prevents the formation of RE clusters,[22] even at low N$_2$ flow (0.75 sccm). Since no RE clustering has been detected for any N$_2$ flow studied, the scenario of an activation of emitting centers (case ($\alpha$)) can be excluded.

For deeper investigation, EDX analysis was performed as a function of depth on the 2 sscm sample for a 700°C annealing (Figure 3d). It was concluded that Ce ions are present across the whole film thickness with an increased signal intensity close to the substrate over a thickness of a few nanometres, meaning a diffusion of Ce ions towards the substrate-film interface upon annealing. The EDX signal detected in the substrate is the background and not related to the presence of Ce ions. This diffusion has already been seen in other systems for high $T_A$ (1200°C) treatments where Ce silicate is formed at the substrate-film interface.[35] But in our case, the analysis on as-deposited layer shows the same Ce distribution profile indicating that this is not due to the thermal budget. This migration is due to the diffusion of Ce into the region near the interface as a way to lower the energy of the system .[22,38] In addition, no Ce oxide (*e.g.* CeO$_2$ or Ce$_6$O$_{11}$)[16,27] or eventually Ce silicates (*e.g.* Ce$_2$Si$_2$O$_7$ or Ce$_{4.667}$(SiO$_4$)$_3$O))[27] have been detected in samples annealed at 900 °C (see Figure 3b) eliminating such luminescence centers discussed previously. Thus, since the BT can be excluded as discussed above, the observed PL band intensity detected at 450 nm for layer deposited under N$_2$ gas flow at 2 sccm seems to be related to the isolated Ce$^{3+}$ ion luminescence centers.

## XPS analysis

XPS was performed to investigate the nature of Ce ions involved, because Cerium ions can exist both under trivalent (Ce$^{3+}$) or tetravalent (Ce$^{4+}$) forms. Indeed, since the target used was CeO$_2$, one can expect that the potential presence of Ce$^{4+}$ ions in the samples could be a barrier for any applications due to its non-optical activity. Thus, it is necessary to determine the Ce oxidation level in the samples which can be determined by XPS measurements. High resolution XPS spectra were fitted for a sputtering time of 240s corresponding to the middle of the film thickness. The O, Si, N and Ce concentrations were 4.9, 62.5, 32.0, 0.6 at.% for 0.75 sccm sample and 11.1, 45.6, 42.7, 0.6 at.% for 2 sccm sample, respectively and are correlated with the RBS measurements (see RBS section - Figure 7).

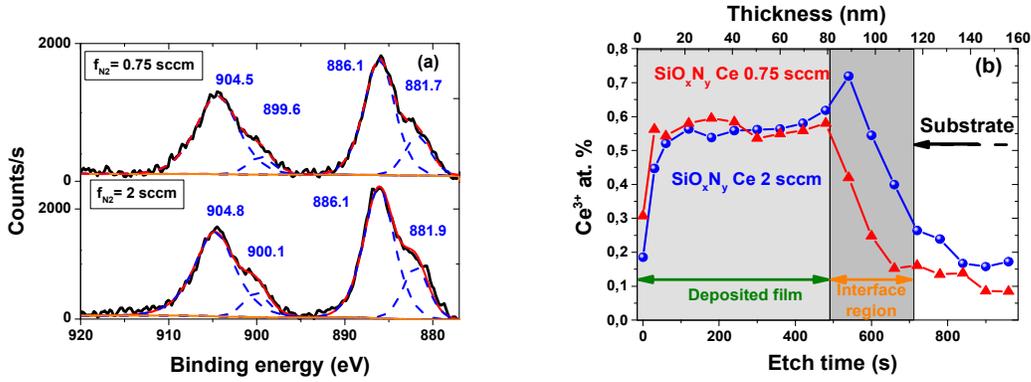

*Figure 4. (a) XPS signature of Ce-3d core level for two $SiO_xN_y$ as-deposited matrices with 0.75 and 2 sccm of $N_2$ flux in the middle of the film, (b) corresponding XPS profiles of Ce-3d distribution*

XPS spectra of the as-deposited films with 0.75 sccm and 2 sccm $N_2$ flow are displayed in Figure 4a. The spectra are very similar for both films. Two large peaks emerge at about 880-890 eV and 900-910 eV fitted with four components given in table 1 and typically attributed to a $Ce^{3+}$ ion signature.

*Table 1: XPS determined Ce-3d peaks attribution, binding energy and separation energy between components and for films grown with different $N_2$ flux rates.*

| Peak Assignment | Binding Energy (eV) | Separation Energy (eV) | Binding Energy (eV) | Separation Energy (eV) | Binding Energy (eV) | Separation Energy (eV) |
|---|---|---|---|---|---|---|
| | 0.75 sccm | | 2 sccm | | [41] (Table 4) | |
| $v_o$ | 881.7 | 4.4 | 881.9 | 4.2 | 881.2 | 4.0 |
| $v'$ | 886.1 | | 886.1 | | 885.2 | |
| $u_o$ | 899.6 | 4.9 | 900.1 | 4.7 | 898.9 | 4.2 |
| $u'$ | 904.5 | | 904.8 | | 903.1 | |

The energy separations of the film grown with 2 sccm are closed to those observed in $Si_3N_4$ matrix (see Ellipsometry section) with $v'-v_o$=4.2eV and $u'-u_o$=4.7 eV. Independently of the charge correction, Anandan et al. found, from $CeO_2$ deposited on $Si_3N_4$ (annealed at 600 °C) the following characteristic differences in the peaks associated with $Ce^{3+}$, $v'-v_o$=4.0 eV and $u'-u_o$=4.2 eV (Table 1).[41] Taking into account that $CeO_2$ on $Si_3N_4$ and Ce doped $Si_3N_4$ are different systems, the energy differences found by XPS are consistent. The XPS characteristic peak at 918 eV, usually attributed to $Ce^{4+}$ ion, is not detectable in this study.[41,42] This also indicates that the cerium electronic state is exclusively $Ce^{3+}$. Figure 4b shows the $Ce^{3+}$ ions distribution in two selected films as a function of depth. In order to explain the obtained results, we have calibrated the experimental data using the relationship between the etching time and layer thickness. Three different zones were identified: the deposited film with thickness ~100 nm, the substrate and an interface region located between the substrate and film with an approximate thickness of 35 nm. Figure 4b shows that in the as-deposited films, $Ce^{3+}$ ions are detected for both films in similar concentrations of 0.5-0.6 at.%. On the other hand, in the interface region under $N_2$ flux rate at 2 sccm, an increase of $Ce^{3+}$ ions concentration is detected indicating the migration of the Ce element toward the substrate-film interface and confirming the conclusion of the TEM analysis presented in Figure 3d *(see TEM section)*. At the same time, the $Ce^{3+}$ migration is less pronounced, and even does not seem to occur, for the sample deposited under lower $N_2$ flux rate of 0.75 sccm. Indeed, due to the lower incorporation, a less strain should be induced in the interface region resulting in a reduced $Ce^{3+}$ ions migration.[22] Note that $Ce^{3+}$ ions appear to be detected in the substrate making think of a diffusion in the substrate. This is only due to the size of the X-ray spot (200 μm) and the lack of lateral resolution of the $Ar^+$ sputtered crater and should not be taken into account.

In summary, the $N_2$ flow seems not to play a role in the changing of the cerium ions oxidation state during the deposition, since $Ce^{4+}$ ions are not present within the films. Thus, the possible emission as suggested by scenario β, involving $Ce^{4+}$ ions correlated with oxygen vacancies in $CeO_2$ to produce a $Ce^{3+}$ emission, is highly unlikely. For both $N_2$ flows, the oxidation state of Ce is fully (III), under which $Ce^{3+}$ ions are optically active allowing a possible Ce activation and PL signal in $SiO_xN_y$ layers. Only the γ hypothesis is able to explain the observed PL emission under high $N_2$ flow which would influence the microstructural environment of the 5d band.

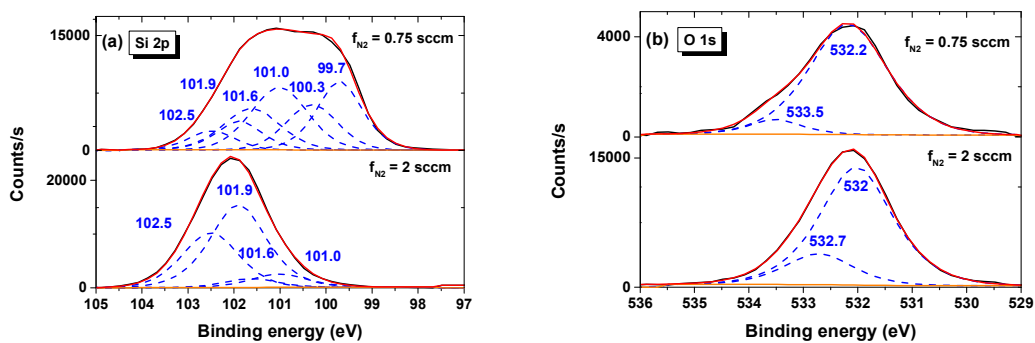

*Figure 5. Fitted XPS spectra (a) Si 2p and (b) O 1s on Ce-doped $SiO_xN_y$ samples with 0.75 sccm and 2 sccm flux rates and annealed at 700°C. The blue values are the peak positions in eV units.*

Figure 5a shows the Si-2p fitted XPS spectra of the two annealed samples fabricated with 0.75 and 2 sccm $N_2$ flows. No analysis on Ce-Si bonding could be done on the Si 2p spectra, because of the very low Ce concentration (*i.e.* 0.6 at.%) with respect to Si (*i.e.* 32-46 at.%). So only the Si chemical environments with O and N major elements can be discussed. Considering these two spectra, significant differences appear. For the 0.75 sccm spectrum, six components were required to obtain a fit, while only four are found for the sample produced with 2 sccm $N_2$ flow. For the latter, the major $Si2p_{3/2}$-$Si2p_{1/2}$ doublet located at 101.9 eV-102.5 eV is typical of a $Si_3N_4$ bonding state,[41,43–45] and is dominant in the spectrum. The doublet peaking at lower energy (101.0 eV-101.6 eV) is attributed to a poorer nitrogen environment for the Si atoms, typically $SiN_x$ (x<4/3) and has minor contribution.[44] Thus, reducing the $N_2$ flow (0.75 sccm) favors the 101.0 eV-101.6 eV doublet while the 101.9 eV-102.5 eV $Si_3N_4$ one decreases compared to the sample grown with 2 sccm flow. Note that the 99.7 eV-100.3 eV doublet typical of a $Si^0$ environment from the elemental Si disappears for the highest $N_2$ flow. No Cerium silicate was observed by TEM and in addition, the $Si^{4+}$ chemical state, generally observed in Cerium silicate (*e.g.* $Ce_{4.67}Si_3O_{13}$, $Ce_2SiO_5$, or $Ce_2Si_2O_7$) at around 103-104 eV, has not been detected either confirming the absence of such Ce clusters[46,47].

Considering the O1s spectra, the major component is located at around 532.1 eV for both samples and is attributed to a $SiO_xN_y$ environment as mentioned in the literature (Figure 5b).[48]

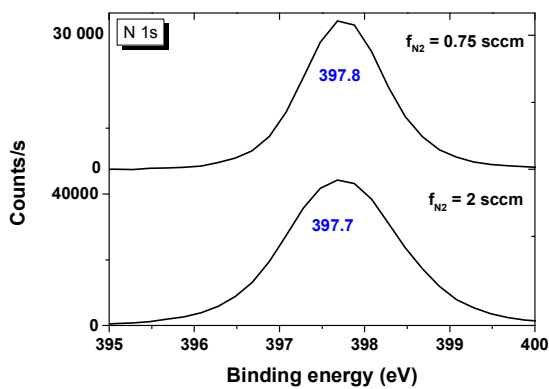

Figure 6. Ns 1s core levels XPS spectra of Ce-doped $SiO_xN_y$ samples fabricated with $N_2$ flow rates values of 0.75 and 2 sccm.

N 1s XPS spectra are displayed in Figure 6 for both $N_2$ flows. The spectra exhibit a major component centered at around 397.8 eV which is typical of a nitride environment and logically attributed to a silicon nitride bonding with the formation of a $SiN_x$ (x<4/3) phase.[44,49,50] However, the full-width half-maximum (FWHM) of the 2 sccm sample is larger (1.63 eV) than that of the 0.75 sccm counterparts (1.21 eV). This enlargement could come from two components, but not enough spectrally split to be well fitted (not shown here). We can assume the presence of $SiO_xN_y$ bondings in the 2sccm sample because of the higher content of O (10 at.% versus 4 at.%).

To conclude, these XPS results show that the cerium element is in $Ce^{3+}$ valence state. A Si-rich $SiN_x$ (x<4/3) matrix is achieved for low $N_2$ flow (0.75 sccm) with free Si atoms, while for higher $N_2$ flow (2 sccm), a matrix with both $Si_3N_4$ and $SiO_xN_y$ bondings without free Si atoms is obtained. To get more information on the different layers, RBS measurements, spectroscopic ellipsometry and FTIR measurements have been carried out and results are presented in the following sections.

## Composition by RBS analysis

Figure 7 displays the RBS measurements of the as-deposited samples with a $N_2$ flow varied from 0.5 sccm to 2 sccm and deposited with a constant RF power density on Si and $CeO_2$ targets fixed at 4.3 W/cm² and 0.4 W/cm², respectively.

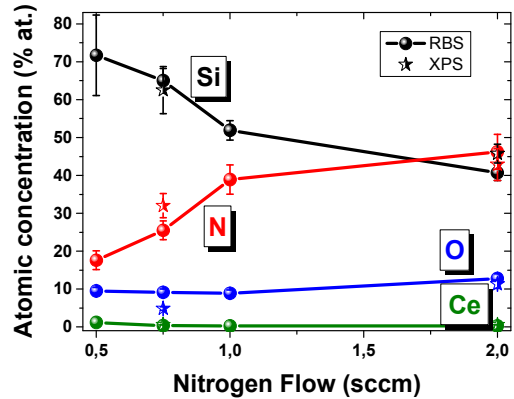

*Figure 7. Si, N, O and Ce atomic concentrations of the Ce-doped SiO$_x$N$_y$ films as a function of N$_2$ flow obtained by RBS analysis (● symbols) and XPS measurements (✶ symbols).*

Low concentrations are measured from 8.9 to 12.8 at.% for oxygen and 1.2 to 0.3 at.% for cerium with respect to N$_2$ flow. This O amount is provided from the CeO$_2$ target. The concentrations deduced from the XPS measurements are consistent with the RBS results. The low discrepancy in Ce concentration values among XPS (0.6 at. % Ce at 2 sccm) and RBS (0.3 at. % Ce at 2 sccm) techniques can be explained by the use of different radiation source and the fact that XPS is a surface sensitive technique. It appears that the N incorporation is done at the expense of the Si. Indeed, the significant concentration increase of N atoms occurs (~16 to 46 at. %) along with a decreasing of Si (~70 to 40 at.%) in the same extend (30%). This same behavior was observed in previous studies in Er-doped silicon-rich oxynitrides (SRON) with the same experimental set up.[51] This was also observed in the case of PECVD technique for the N and O incorporations, which substitute to Si atoms in order to form SiO$_x$N$_y$.[52] The composition found with RBS measurements are in the same tendency as in XPS analysis (see Figure 7).

**Ellipsometry analysis**

The ellipsometry technique allows the determination of the optical parameters of the films such as the sub-layer thicknesses and the wavelength dependence of the optical index of each involved material. In this regard, we have conducted ellipsometry analysis for the as-deposited samples. The real part of the refractive index versus the wavelength is plotted in Figure 8a for several as-deposited samples grown with different N$_2$ fluxes (solid lines). Those curves were compared to optical indexes curves for the Si, Si$_3$N$_4$ and SiO$_2$ phases (full lines). As expected, the Si peak centered at 375 nm is not observed for any of the as deposited layers, indicating that there is a mix of elements in the same phase instead of highlighted phase separation (*i.e.* Si, Si$_3$N$_4$ and SiO$_2$). The refractive index (left scale) and layer thickness (right scale) are displayed as function of N$_2$ flow (see Figure 8b), with horizontal dashed lines depicting the Si, Si$_3$N$_4$ and SiO$_2$ refractive indexes. For samples grown at low N$_2$ flow (0.5 sccm), the refractive index is equal to 3.8 at 1.95 eV (636nm) which corresponds to Si, confirming the low N$_2$ incorporation in the layer. Increasing N$_2$ flow decreases the refractive index of the layers down to values corresponding to the refractive index of stoichiometric Si$_3$N$_4$ ($n_{Si3N4} \approx 2.0$). For the samples grown at higher N$_2$ flow (2 sccm), the refractive index is even slightly below 2.0 (n≈1.9) which corresponds to $SiO_xN_y$ matrix with low oxygen concentration.[53] Concerning the layers thicknesses, a decrease of almost 30 % is observed, when the $N_2$ flow changes. Indeed, the mean free path of particles in the plasma decreases with increasing $N_2$ flow so that the deposition rate is reduced leading to a decrease of the thickness of the films.[22,54]

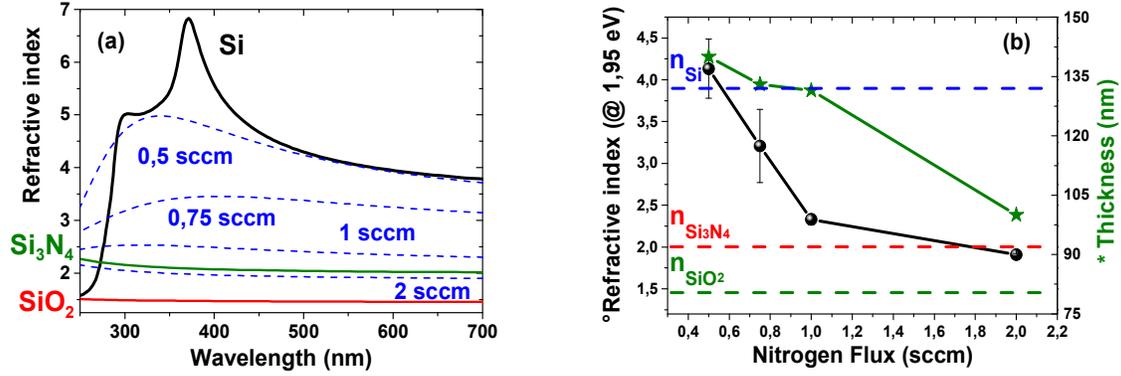

Figure 8. (a) Refractive index as a function of the wavelength according to the N fluxes and different stoichiometric phases (b) Refractive index and thickness evolution of the as-deposited Ce-doped $SiO_xN_y$ films as a function of $N_2$ flow.

**Bruggeman model**

From the ellipsometry data, a modeling of the material taking into account 3 phases has been carried out by using the effective medium theory (EMT).[53,55] The modeling took into account only the 500-700 nm part of the measured spectrum since the experimental data do not show the Si peak at 375 nm. The EMT theory is described by the equation (1):

$$V_{Si}\left(\frac{\varepsilon_{Si} - \varepsilon_h}{\varepsilon_{Si} + 2\varepsilon_h}\right) + V_{SiO2}\left(\frac{\varepsilon_{SiO2} - \varepsilon_h}{\varepsilon_{SiO2} + 2\varepsilon_h}\right) + V_{Si3N4}\left(\frac{\varepsilon_{Si3N4} - \varepsilon_h}{\varepsilon_{Si3N4} + 2\varepsilon_h}\right) = \frac{\varepsilon - \varepsilon_h}{\varepsilon + 2\varepsilon_h} \quad (1)$$

where $V_{Si}$, $V_{SiO2}$, $V_{Si3N4}$ are the volume fractions of Si, $SiO_2$ and $Si_3N_4$ phases respectively, and $\varepsilon_{Si}$, $\varepsilon_{SiO2}$ and $\varepsilon_{Si3N4}$ the dielectric permittivities of each phase, $\varepsilon_h$ is the dielectric permittivity of the host matrix and $\varepsilon$ is the one measured by ellipsometry. In the Bruggeman effective medium approximation, $\varepsilon$ is taken equal to $\varepsilon_h$ and the layer is considered as a random mixing of three phases (here Si, $Si_3N_4$ and $SiO_2$). The sum of the three individual volume fractions is equal to 1. This approach results in the following equations (2) and (3):

$$V_{Si} + V_{Si3N4} + V_{SiO2} = 1 \quad (2)$$

$$V_{Si} \cdot \left[\left(\frac{\varepsilon_{Si} - \varepsilon_h}{\varepsilon_{Si} + 2\varepsilon_h}\right) - \left(\frac{\varepsilon_{SiO2} - \varepsilon_h}{\varepsilon_{SiO2} + 2\varepsilon_h}\right)\right] + V_{Si3N4} \cdot \left[\left(\frac{\varepsilon_{Si3N4} - \varepsilon_h}{\varepsilon_{Si3N4} + 2\varepsilon_h}\right) - \left(\frac{\varepsilon_{SiO2} - \varepsilon_h}{\varepsilon_{SiO2} + 2\varepsilon_h}\right)\right] + \left(\frac{\varepsilon_{SiO2} - \varepsilon_h}{\varepsilon_{SiO2} + 2\varepsilon_h}\right) = 0 \quad (3)$$

*Table 2. Calculated volume fractions of different phases as function of $N_2$ flow predicted by the Bruggeman model.*

| Nitrogen Flux (sccm) | | 0.50 | 0.75 | 1.00 | 2.00 |
|---|---|---|---|---|---|
| Volume fractions | $V_{Si}$ | 0.88 | 0.50 | 0.19 | 0.03 |
|  | $V_{Si3N4}$ | 0.01 | 0.35 | 0.65 | 0.70 |
|  | $V_{SiO2}$ | 0.11 | 0.15 | 0.16 | 0.27 |
| $\chi^2$ | | 0.13 | 0.09 | 0.04 | 0.004 |

The phases volume fractions have been deduced from the least squares method and displayed in Table 2 with the $\varepsilon_h$ deduced from the refractive index of Figure 8a for each $N_2$ flow, while $\varepsilon_{Si}$, $\varepsilon_{SiO2}$ and $\varepsilon_{Si3N4}$ were taken from database.[56] The O concentration was taken from RBS measurements to have an approximation of the $V_{SiO2}$ phase and was injected into the equations passing from an equation with 3 unknowns to 2 unknowns. The data in Table 2 shows that a low $N_2$ flow (0.5 sccm) leads to a huge Si content, as expected, while for the intermediate $N_2$ flows (0.75 sccm and 1 sccm) the volume fractions of the Si and $Si_3N_4$ phases reverse. Finally, $Si_3N_4$ phase is predominant for a high $N_2$ flow (2 sccm), with an increase of the $SiO_2$ volume fraction linked to the important Si phase reduction, that explains the achievement of a refractive index below $Si_3N_4$ one (see Figure 8b).

As seen above (see TEM section), although the microscopy analysis does not reveal a phase separation for any $T_A$, the Bruggeman model can give the behavior of the atomic concentration through the phase proportion (Eq.3). The atomic fractions ($f_i$) of each species $i$ ($i$=Si, $Si_3N_4$ and $SiO_2$) are deduced from the phase volume fractions, with help of equations 4, 5 and 6, in order to obtain the atomic fraction of Si, N and O, respectively:

$$f_{Si} = V_{Si} \cdot \frac{\rho_{Si}}{M_{Si}} + 3 \cdot V_{Si3N4} \cdot \frac{\rho_{Si3N4}}{M_{Si3N4}} + V_{SiO2} \cdot \frac{\rho_{SiO2}}{M_{SiO2}} \quad (4)$$

$$f_N = 4 \cdot V_{Si3N4} \cdot \frac{\rho_{Si3N4}}{M_{Si3N4}} \quad (5)$$

$$f_O = 1 - (f_{Si} + f_N) \quad (6)$$

where $\rho_i$ and $M_i$ are the density and the molar weight of the phase $i$.[57]

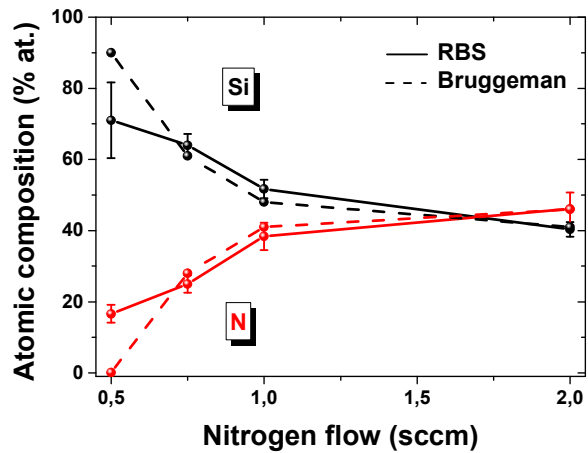

*Figure 9. Comparison of Si and N atomic concentrations in Ce-doped $SiO_xN_y$ films deduced from RBS measurements and Bruggeman modeling.*

Figure 9 shows a comparison of the atomic concentrations obtained by RBS measurements and by Bruggeman modeling. The N and Si atomic concentrations obtained by Bruggeman modeling are close to the ones deduced from the RBS experiments for $N_2$ flux going from 0.75 up to 2 sccm flux. For the 0.5 sccm flux, a larger difference is observed which can be explained by the uncertainty of the measured $\varepsilon$ parameter. Indeed, for the low N flux, the layer is mainly composed of silicon deposited on a Si substrate. The ellipsometry measurements are not precise enough because of the weak index contrast between the Si substrate and the subsequent film. Finally, the modeling demonstrates that with a $N_2$ flux below 2 sccm, a Si "phase" is present and likely kills the expected luminescence of Ce ions, as observed previously for enriched Ce-doped $SiO_{1.5}$ films on Si.[40] The $\gamma$ scenario mentioned above in the optical analysis part, involving the surrounding media of Ce, is then confirmed. For 2 sccm, the samples contain mainly two phases ($Si_3N_4$ and $SiO_2$) which exhibit a Ce PL signal since no free Si phase is detected.

## Fourier transform infrared spectroscopy

FTIR analysis was performed to characterize the influence of $N_2$ flux on the Si-N bonds, as well as the Si-O bonds resulting from the O atoms supplied by the sputtering of the $CeO_2$ target. Figure 10 displays FTIR spectra measured at Brewster angle of as-deposited samples with different $N_2$ flows.

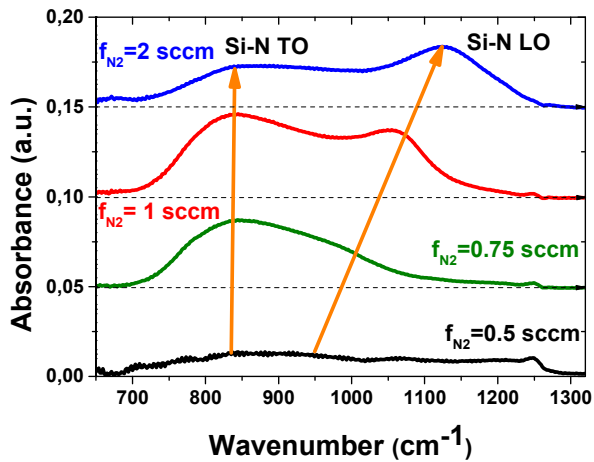

*Figure 10. Infrared spectra (FTIR) of the Ce-doped SiO$_x$N$_y$ films depending on nitrogen flow.*

For the sample grown at the lowest N$_2$ flow (0.5 sccm), an almost flat spectrum is observed without sharp peak. The N$_2$ flux favors the N incorporation as evidenced by the appearance of two vibration bands around 840 and 990 cm$^{-1}$ for 0.75 sccm. Those two peaks correspond to *TO$_{Si-N}$* transversal optical mode and *LO$_{Si-N}$* longitudinal optical mode[58] and shift towards high energies with increasing flux. The *LO$_{Si-N}$* band blueshifts from 1000 cm$^{-1}$ for 0.75 sccm to 1060 cm$^{-1}$ for samples grown at 1 sccm and 1125 cm$^{-1}$ for the 2 sccm, respectively, while the *TO$_{Si-N}$* band shifts in a lesser extent. Those band shifts have already been observed in SiN$_x$:H films and SiN$_x$ films. In SiN$_x$:H films, the shift was attributed to the replacement of a N atom by a H atom in a Si-N bond.[59] In SiN$_x$ films, Debieu and al.[58] observed this same blueshift and noted also that the *TO$_{Si-N}$* band position is less sensitive to the composition. The origin of this blueshift is attributed to a better organization of the layer. Indeed, an N incorporation tends to stabilize Si$_3$N$_4$ phase as mentioned in Table 2.

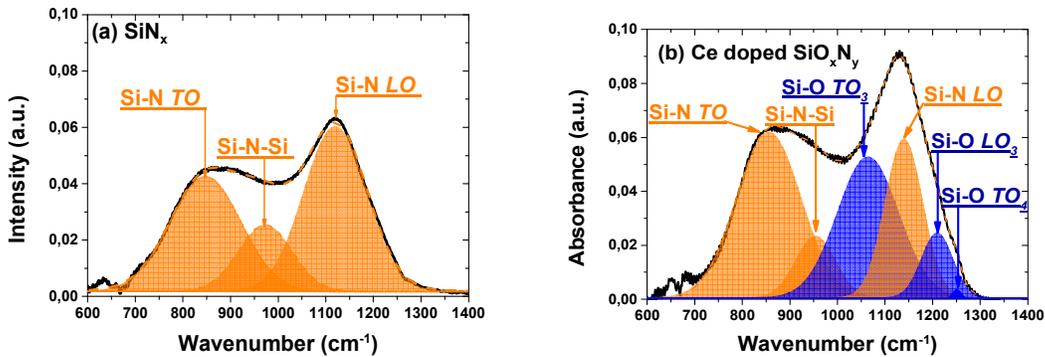

*Figure 11. (a) Infrared (FTIR) spectrum of SiN$_x$ film for a 2sccm N$_2$ flux, (b) infrared spectrum of Ce doped SiO$_x$N$_y$ sample deposited with 2 sccm N$_2$ flux.*

Among the different N$_2$ fluxes, a PL emission is only observed for the 2 sccm sample (Figure 1a) which is then investigated hereafter. To identify the Si-N peaks in the 2 sccm sample, a SiN$_x$ reference sample with the same $N_2$ flow rate at 2 sccm but without Ce and consequently without O doping has been deposited. It was found a refractive index close to $Si_3N_4$. Figure 11a shows the FTIR spectrum along with the fit revealing three peaks centered at 848, 974 and 1121 cm$^{-1}$. Such peaks were identified in previous studies on $SiN_x$ matrices [22,23,60] and correspond to the $TO_{SiN}$, ≡ Si-N-Si ≡ bond and $LO_{SiN}$ mode, respectively. With help of this first identification step on Si-N bonds, the FTIR spectrum of the Ce and O doping sample grown with 2 sccm $N_2$ flow has been displayed in Figure 11b. Three new peaks are detected in the fitting process and attributed to the Si-O bonds in addition to the three Si-N bonds observed in $Si_3N_4$ matrix. The Si-N bonds are centered at 856, 955 and 1140 cm$^{-1}$, close to the $Si_3N_4$ peak positions aforementioned above. Two peaks centered at 1065 and 1211 cm$^{-1}$ are ascribed to the $TO_{3\ Si-O}$ transversal optical mode and $LO_{3\ Si-O}$ longitudinal optical modes respectively confirming the oxygen incorporation in the layer. [61–65] The last peak at 1250 cm$^{-1}$ corresponds to the Si-O $TO_4$ modes.[66]

Therefore, the presented FTIR data and analysis confirm the existence of a phase similar to the expected $Si_3N_4$ phase, which support further the claim that synthesized Ce-doped $SiO_xN_y$ material has a good organization.

## Excitation mechanism

The discussion presented in the PL and PLE section, highlighted different assumptions and scenarios related to the nature of emitting centers and how they are respectively activated. Specifically, the (i) band tails, (ii) $CeO_2$ and (iii) Ce clusters were excluded in favor of the (iv) presence of $Ce^{3+}$ ions (see XPS section) and their signature in the PLE spectrum (see PL and PLE section). In the same way, the different scenarios were proposed to explain the $Ce^{3+}$ emission for only a high $N_2$ flux (2 sccm). Namely, we have considered three cases: ($\alpha$) the decrease of Ce clusters formation with the $N_2$ flux was mentioned as a possible cause; however no cluster was observed for any $N_2$ flux investigated, ($\beta$) a possible modification of the oxidation state from Ce (IV) to (III) was also cited, but rejected by cross-checking the PL and XPS results. ($\gamma$) Finally, the surrounding host structure of Ce was considered. In this scenario, two different mechanisms were proposed: the nephelauxetic effect and Si atoms surrounding Ce that annihilate the emission. These last two mechanisms are discussed in details below.

Concerning the nephelauxetic effect, the $Ce^{3+}$ ion emission had been reported at different emission wavelengths depending on the host matrices. Usually, this emission is located in the UV-blue spectral region but could be observed also in the green or red regions for $Ce^{3+}$ doped $YSi_xO_yN_z$ or $Y_2O_2S$ matrices, respectively.[67,68] For our $SiO_xN_y$ matrix, a previous study showed a red shift of $Ce^{3+}$ ion emission from 400 nm for N-free samples to 476 nm for high N content (40 at. %).[20] This PL shift, called the nephelauxetic effect, comes from the interaction between the 5d band and the local environment.[69] Note that this effect is amplified by the fact that the N content is predominant compared to O as underlined by Y. Li *et al*.[36,37] Such an effect is illustrated in Figures 12a and 12b for the extreme limits of the N content, respectively.

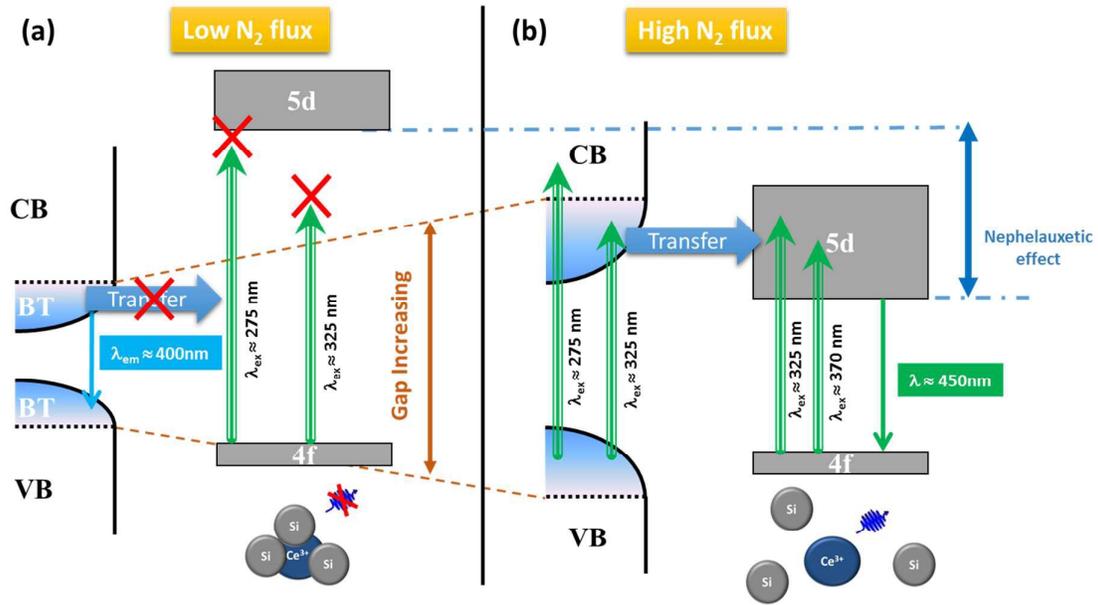

*Figure 12. Schematic representation of the nephelauxetic effect (scenario γ), the Si atoms surrounding Ce and the band gap variation for (a) low and (b) high nitrogen flux.*

In our study, for samples grown at low $N_2$ flux (0.5 to 1 sccm), the 5d band could be localized at higher energy owing to the nephelauxetic effect (lower wavelength 325 nm) (see Figure 12a). No $Ce^{3+}$ ion emission was detected at 325 nm excitation wavelength (Figure 1a) and only the BT emission centered at 400 nm is detectable. If the shift caused by the nephelauxetic effect is so large then such excitation energy is not enough to excite the 5d band. Interestingly enough, even for higher excitation (275 nm) no emission is observed (not shown here), meaning that the nephelauxetic effect is not responsible for the $Ce^{3+}$ ion optical activation. In this case, Si atoms surrounding $Ce^{3+}$ ions blocks the $Ce^{3+}$ ion emission. Indeed, the previous results shown that a higher Si concentration for lower $N_2$ flux allow $Ce^{3+}$ ions to back transfer their energy to neighbor Si atoms. Hence, such an environment kills the $Ce^{3+}$ ion emission.[70]

A last point needs more clarification. Even if using a RE like cerium is advantageous considering the $Ce^{3+}$ ion direct excitation process present in our system as mentioned at the beginning of this study, the matrix absorption cannot be ignored nor the possible matrix-RE energy transfer. Due to the fact that N incorporation to the host matrix could be controlled, the matrix gap changes from a low value around 1.1 eV for the lowest N content (close to Si bulk) and tends to become similar to the $Si_3N_4$ bandgap of 3.3 eV for higher $N_2$ flux. The bandgap increases even further (~ 4.0 eV) for the samples grown with $N_2$ flux at 2 sccm corresponding to $SiO_xN_y$. Therefore, due to the simultaneous crossing of the increasing bandgap and the nephelauxetic effect (*i.e.* lower energetic position of the 5d level),[71,72] one can expect an overlapping effect as schematically shown in Figure 12b. To have a deeper understanding of this overlap, the PLE spectrum and the corresponding PL spectra at different excitation wavelengths of the 2 sccm sample are shown in Figures 13a and 13b, respectively. The PLE spectrum is the same as Figure 1b, which indicated a direct excitation of the $Ce^{3+}$,[19,32] but now in a larger excitation range from 250 nm to 400 nm. In this range, the PL spectra (Figure 13b) have the same shape as the $Ce^{3+}$ ion emission seen in Figure 1a (excited at 325 nm ), whatever the excitation wavelengths between 250 nm and 400 nm. This fact demonstrates that $Ce^{3+}$ ion can effectively excite in a large spectral range.

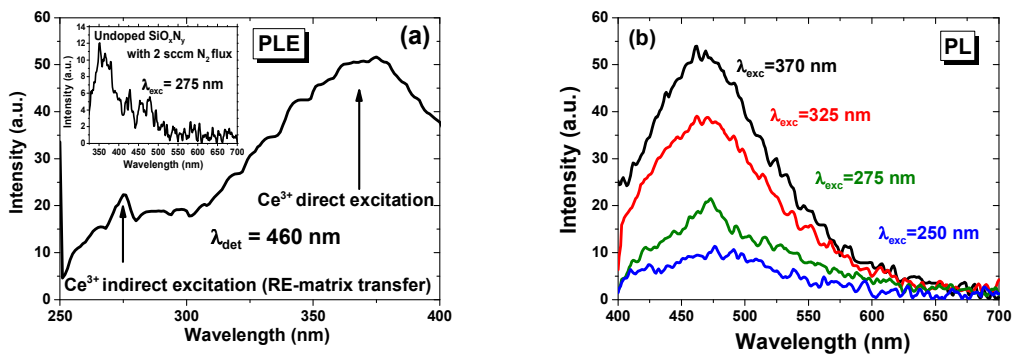

*Figure 13. (a) PLE spectrum of Ce-doped $SiO_xN_y$ measured in 250-400 nm spectral range and (b) the corresponding PL spectra of $Ce^{3+}$ ion in $SiO_xN_y$ samples deposited with a 2 sccm $N_2$ flux annealed at 700 °C. The inset in panel (a) shows the PLE spectrum of reference undoped $SiO_xN_y$ sample.*

Then, either the 5d energy band is much enlarged at high $N_2$ flux making direct excitation easy or the matrix comes into play and favors an indirect excitation mechanism. Indeed, the PL of the undoped $SiO_xN_y$, as reference sample, at high $N_2$ flux (inset of Figure 13a), shows a maximum emission at 350 nm when excited at 275 nm. This emission wavelength corresponds to the maximum excitation range detected in the $Ce^{3+}$ PLE spectrum (Figure 13a). Such an overlap confirms an indirect energy transfer from the matrix to the $Ce^{3+}$ ion (Figure 12b), for the shorter excitation wavelength (Figure 13a). In conclusion, a direct excitation of $Ce^{3+}$ ions is predominant between 300 and 400 nm and an indirect excitation, from 250 nm to 275 nm at least, is possible via the matrix with the high $N_2$ flux sample (Figure 13a), whereas such excitations mechanism are not possible for samples grown with low $N_2$ flux rates (0.5 to 1 sccm).

## Evolution of the photoluminescence with higher Ce concentration

Taking into account all the growth parameters already discussed for achieving a good $Ce^{3+}$ ion emission *(i.e.* 2 sccm $N_2$ flux), a new set of Ce-doped samples deposited with higher RF power density applied on $CeO_2$ target (up to 2.1 W/cm²) have been deposited. Specifically, this set includes the undoped $SiO_xN_y$ reference sample (see inset of Figure 13a), the best $Ce^{3+}$ ion emitting thin film (see Figure 1) containing 0.3 at.% of Ce measured by RBS and two additional $Ce^{3+}$ ion samples having 4 at.% and 6 at.% of Ce measured also by RBS. Figure 14a displays the PL spectra and Figure 14b shows the corresponding integrated PL intensity of these samples as a function of the RF power density applied and the $Ce^{3+}$ concentration (integrated PL intensity was obtained by integrating the PL spectra between 400 and 675 nm).

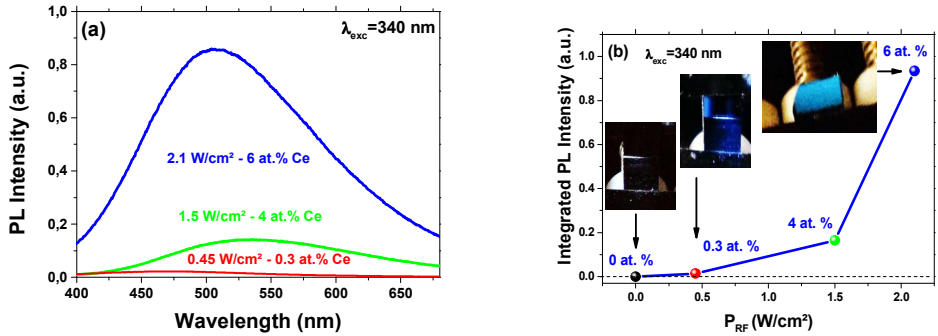

*Figure 14. (a) PL spectra of $SiO_xN_y$: $Ce^{3+}$ films deposited with different power densities applied on $CeO_2$ target and (b) the corresponding integrated PL intensity. Insets show optical images of the $SiO_xN_y$: $Ce^{3+}$ samples under UV excitation at 340 nm and their corresponding RBS concentrations*

The PL intensity increases nonlinearly with the RF power density up to 2.1 W/cm² on $CeO_2$ target corresponding to 6 at. %. This increasing of PL intensity attest a good incorporation of $Ce^{3+}$ ions and an efficient optical activity even at high doping. Indeed, the PL intensity is almost 70 times higher at 6 at.% compared to the 0.3 at.% film, and this important $Ce^{3+}$ concentration is 6 times higher compares to other studies.[20]

A similar PL intensity dependence on the dopant content was already observed in other matrices such as $SiO_2$[35,73], but a PL intensity saturation effect is detected due the formation of Ce clusters or silicates. In the present study, the integrated PL intensity shows monotonic increase with the rare earth doping level change, without saturation, confirming the lack of Ce cluster or silicate phase formation due do the presence of nitrogen. Furthermore, we have performed microscopy analysis (not shown here) which confirms that our films are amorphous without any evidence of crystalized precipitates even at the highest Ce content studied (6 at.%). Finally, to demonstrate that developed $SiO_xN_y$: $Ce^{3+}$ thin films are efficient emitters, we took pictures of selected samples excited at 340 nm at room temperature (see Figure 14(b) insets). It can be seen that bright blue emission was observed under the regular room ambient lighting conditions, attesting to a good quantum yield of $SiO_xN_y$: $Ce^{3+}$ with 6 at.% of Ce as compare to the 0.3 at.%. Under the same excitation/detection conditions, the undoped sample does not show observable visually emission.

## Conclusions

In this study, we investigated the influence of reactive nitrogen flux during the deposition of Ce-doped $SiO_xN_y$ films on the matrix structure and the emission behavior of highly doped films.

By increasing the $N_2$ flux, four different matrices were deposited. The first one, at low $N_2$ flux is composed mainly of bulk Si which is confirmed by the high Si concentration revealed by RBS. The fourth, at high $N_2$ flow (2sccm), is a mixing of $SiO_2$ and $Si_3N_4$ phases. The Si, $SiO_2$, $Si_3N_4$ phase volume fractions have been deduced from the Bruggeman modeling. FTIR data confirm the presence of $Si_3N_4$ phases and characteristic interatomic bonds (Si-N and Si-O), of the $SiO_xN_y$ matrix. Above all, the N incorporation in the deposited film is well established during the process. The TEM images don't show clusters or silicates after high temperature annealing, demonstrating the N incorporation effect.

Photoluminescence measurements show a wide blue emission from 400 nm to 650 nm under UV excitation for samples grown at high nitrogen flux (2 sccm) and no emission for lower flux. After considering the possible luminescence centers including band tails, $CeO_2$, Ce clusters, $Ce^{3+}$ ions, the $Ce^{3+}$ ion is identified by XPS analysis and PLE spectroscopy as the origin of the observed emission. The different scenarios relevant to activate the $Ce^{3+}$ emission, according to the N incorporation, are also investigated. All the results converge to indicate that the change of $Ce^{3+}$ ion environment explains the modification of the luminescence properties when free from Si phase. It was concluded that a Si phase is present only in the matrix for materials grown at low N content where it interacts with $Ce^{3+}$ ions causing a non-radiative energy transfer between $Ce^{3+}$ ion and its surrounding host structure. Furthermore, nitrogen induces a red shift of the $Ce^{3+}$ ion emission as compared to Ce-doped $SiO_x$ due to the nephelauxetic effect. In addition, direct and indirect excitations have been highlighted for $Ce^{3+}$ ions in $SiO_xN_y$ host matrix: a direct excitation under 300 to 400 nm wavelength excitation and an indirect excitation via matrix-RE energy transfer is possible for lower wavelength excitation.

Taking into account the optimized nitrogen growth parameters, Ce concentration was increased in the films giving a strong blue emission visible to the naked eye. The integrated intensity of samples doped up to 6 at. % shows no signs of quenching. Moreover heavily doped films are still amorphous and show no traces of Ce clusters or silicates formation. Ce doped silicon oxynitrides are promising donor materials in down-conversion (*e.g.* $Ce^{3+}$-$Yb^{3+}$) as well as light emitting device for electroluminescence due to the bright photoluminescence from $Ce^{3+}$ ion and its high solubility in $SiO_xN_y$ matrix. Such a material system is advantageous for these applications as compared with other Si based matrices where clustering and phase separation are observed.

**Acknowledgements:**

The authors want to thank the "Agence Nationale de la Recherche » (ANR-11-EQPX-0020 project) in the framework of the "Investissements d'avenir" program for financial support (FIB system).

The authors would like to thank Mr Cédric Frilay from CIMAP Laboratory (Caen, France) for his great help on samples growth.

This work was financially supported by the French Research National Agency through the GENESE project (N° ANR-13-BS09-0020-01) .

W.M.J. acknowledges the support from the NSF CAREER Award no. DMR- 1056493.